%% file: main.tex
\documentclass[%
 aip,
 amsmath,amssymb,
 reprint,%
]{revtex4-1}

\usepackage{dcolumn,import}
\usepackage{transparent, color}
\usepackage{appendix}
\usepackage{upgreek}
\usepackage[mathcal]{euscript}
\usepackage{float}

\usepackage{qcircuit}
\usepackage{braket}

\usepackage{graphicx}
\usepackage{bm}

\usepackage[utf8]{inputenc}
\usepackage[T1]{fontenc}
\usepackage{mathptmx}
\usepackage{etoolbox}

\makeatletter
\def\@email#1#2{%
 \endgroup
 \patchcmd{\titleblock@produce}
  {\frontmatter@RRAPformat}
  {\frontmatter@RRAPformat{\produce@RRAP{*#1\href{mailto:#2}{#2}}}\frontmatter@RRAPformat}
  {}{}
}%
\makeatother
\begin{document}

\preprint{AIP/123-QED}

\title{Coplanar waveguide ground potentials imbalance as a source of useful signal in Near-Field Scanning Microwave Microscopy}
\affiliation{National~University~of~Science~and~Technology~MISIS,~119049~Moscow,~Russia}
\author{P.A.~Gladilovich\(^*\)}
    \email{m2003860@edu.misis.ru}

\author{A.V.~Sabluk}

\author{P.S.~Burtsev}

\author{R.~Migdisov}

\author{N.~Maleeva}

\author{S.V.~Shitov}

\begin{abstract}
Various techniques are available in order to obtain information on samples of a different nature in near-field scanning microwave microscopy (NSMM), with transmission-line resonator (TLR) techniques considered as the most advanced in terms of sensitivity and resolution. In this paper, we focus on development of a TLR-based NSMM supplied by a new source of useful signal: coplanar waveguide ground potentials imbalance. Electromagnetic modeling of the device and experimental scanning of two planar structures are conducted to examine the performance of the proposed technique. Both modeling and experimental results  demonstrate the ability to distinguish symmetric inhomogeneity positions with respect to the central conductor of the coplanar waveguide by the phase of imbalance signal. 
The thin-film structure scanning procedure displays the possibility of proposed approach to resolve low-contrast structures with accuracy up to 98.7 \%.
\end{abstract}

\maketitle
\section{Introduction}

Over the past few years, near-field scanning microwave microscopy (NSMM) has been considered as one of the most important methods for quantitative determining the electrodynamic characteristics of materials in the microwave range \cite{Knoll1997,Abu2001, KimKim2003, Gao1998, Fragola2004, Imtiaz2007}. Such microscope has nondestructive and localized access to the distribution of electromagnetic parameters of the material as well as to the structure of the surface and subsurface \cite{Chisum2009, Ren2011, Plassard2011}. NSMM has various applications, including research on doping and defects in nanowires \cite{Berweger2020, Wallis2011}, superconducting materials \cite{Takeuchi1997}, the domain structure of magnetics \cite{Melikyan2009}, and the dopant profiles of semiconductors \cite{Imtiaz2014, Kundhikanjana2011}. Moreover, this kind of microscopy is used to study the conductive properties of photovoltaic materials \cite{Coakley2019, Hovsepyan2009, Weber2012} and to image organics at the cellular level, including the structure of DNA molecules \cite{Farina2012, Lee2013, Wu2015}.

It's common to distinguish several techniques in NSMM \cite{Anlage2001}: (a) microwave cavity resonator with an evanescent hole in one wall \cite{Frait1959, Soohoo1962, Lofland1995, Ikeya1987}, (b) non-resonant microwave transmission line microscope \cite{Bryant1965, Xu1992, Stuchly1980, Burdette1980, Fee1989, Stranick1993, Jiang1993, Asami1994, Gutmann1987, Qaddoumi1997}, (c) SQUID-based microwave microscope \cite{Black1995}, (d) AFM-based near-field scanning microwave microscope \cite{Manassen1994, Zhang1996, Zhang1998, Wago1998}, (e) transmission-line resonator with an open end as an area of sensitivity \cite{Bosisio1970}. Transmission-line resonator technique is considered to be one of the most sensitive forms of NSMM \cite{Anlage2001}. A resonant frequency shift or a quality factor change are used for detecting sample inhomogeneities.

Here we report a NSMM platform demonstrator based on such a conventional technique as transmission-line resonator, but supplemented with one more source of useful signal: coplanar waveguide ground potentials imbalance \cite{Shitov2019}.

\begin{figure}[tbhp]
\centering
\normalsize
 \def\svgwidth{0.48\textwidth}  
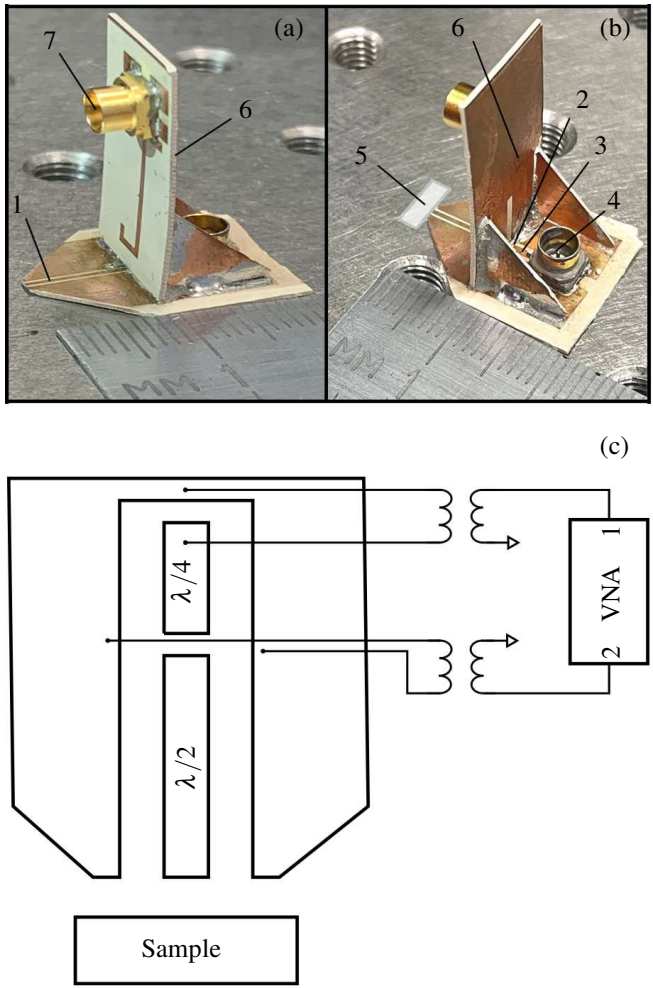
 \newline
\caption{\textbf{Details on the detector.} Front \textbf{(a)} and back \textbf{(b)} views of the near-field scanning microwave detector based on a coplanar half-wave resonator \textbf{1}, connected via capacitive slot \textbf{2} to a signal supplying system, which consists of a coplanar quarter-wave resonator \textbf{3}, acting as a continuation of a coaxial line attached to the detector by an SMP connector \textbf{4}. The area of sensitivity \textbf{5} in near-field regime lies in the vicinity of a half-wave resonator's open end. A bridge structure \textbf{6} is used to collect ground potentials imbalance signal through an SMP connector \textbf{7}. The schematic representation \textbf{(c)} shows the principal elements of the detector: a coplanar half-wave resonator coupled via capacitive slot to a a coplanar quarter-wave resonator (all surrounded by ground) and  VNA used for collecting S-parameters.}
\label{detector_photo}
 \end{figure}

\begin{figure*}[tbhp]
\centering
\normalsize
 \def\svgwidth{\textwidth}  
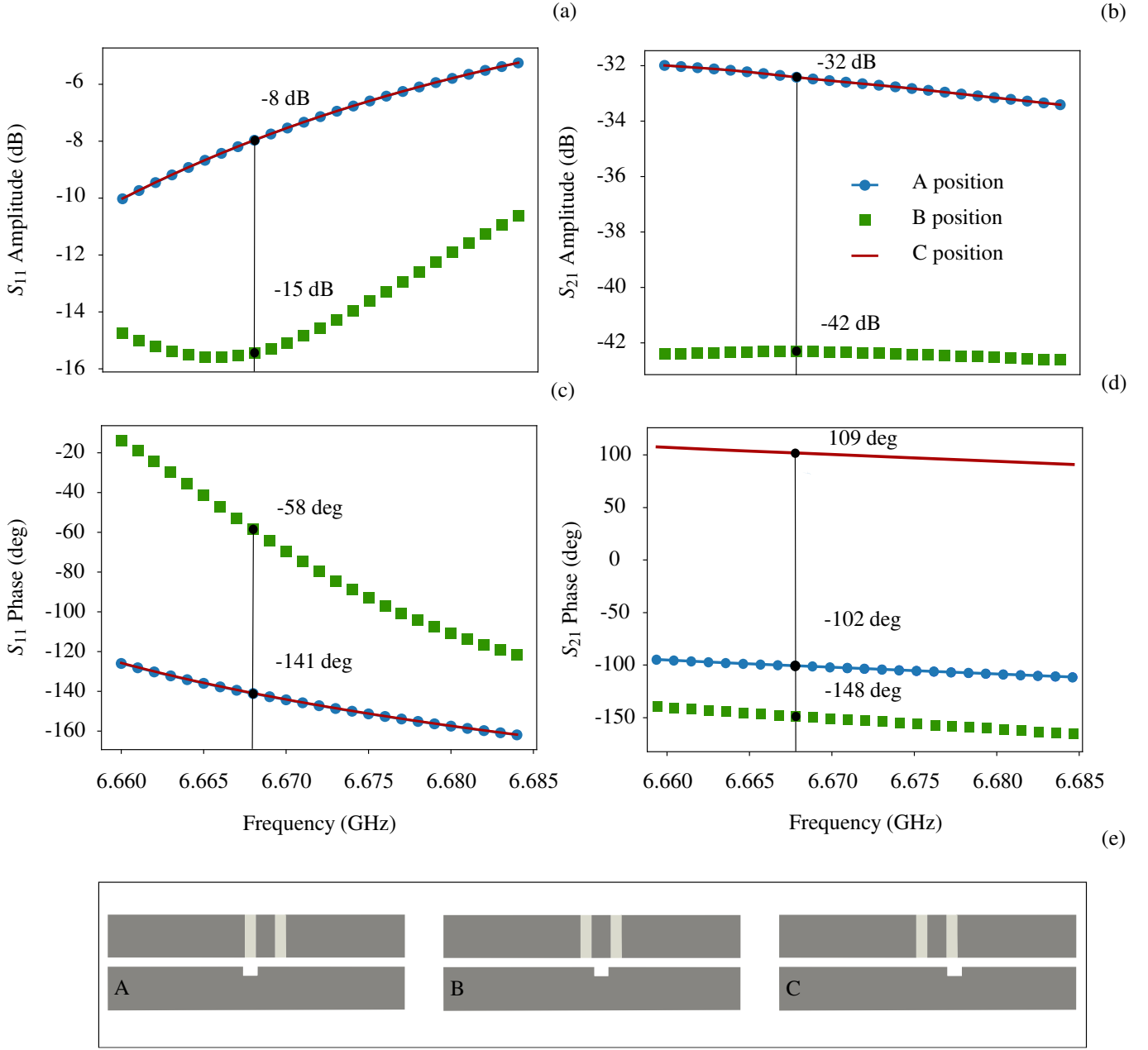
 \newline
\caption{\textbf{Reflection and transmission coefficients simulated for three positions of the recess in a sample.} Amplitudes \textbf{(a,b)} and phases \textbf{(c,d)} of reflection $S_{11}$ and transmission $S_{21}$ coefficients were numerically simulated for three positions of the recess in the cooper plate \textbf{(e)} in order to verify detectors's ability to distinguish positions \textbf{A}, \textbf{B} and \textbf{C}. Positions \textbf{A} and \textbf{C} create same response in both amplitudes and $S_{11}$ phase, while creating approximately $\pi$ difference in $S_{21}$ phase. Vertical line corresponds to a frequency of maximum difference in signals  for central position \textbf{B} and shifted positions \textbf{A}, \textbf{C}: 7 dB at - 15 dBm in $S_{11}$ amplitude; 10 dB at - 42 dBm in $S_{21}$ amplitude. $S_{11}$ amplitude is more useful in order to distinguish central position \textbf{B} from shifted positions \textbf{A} and \textbf{C}, although $S_{21}$ phase is good for distinguishing positions \textbf{A} and \textbf{C}}
\label{ABC_responce}
\end{figure*}

\begin{figure*}[tbhp]
\centering
\normalsize
 \def\svgwidth{\textwidth}  
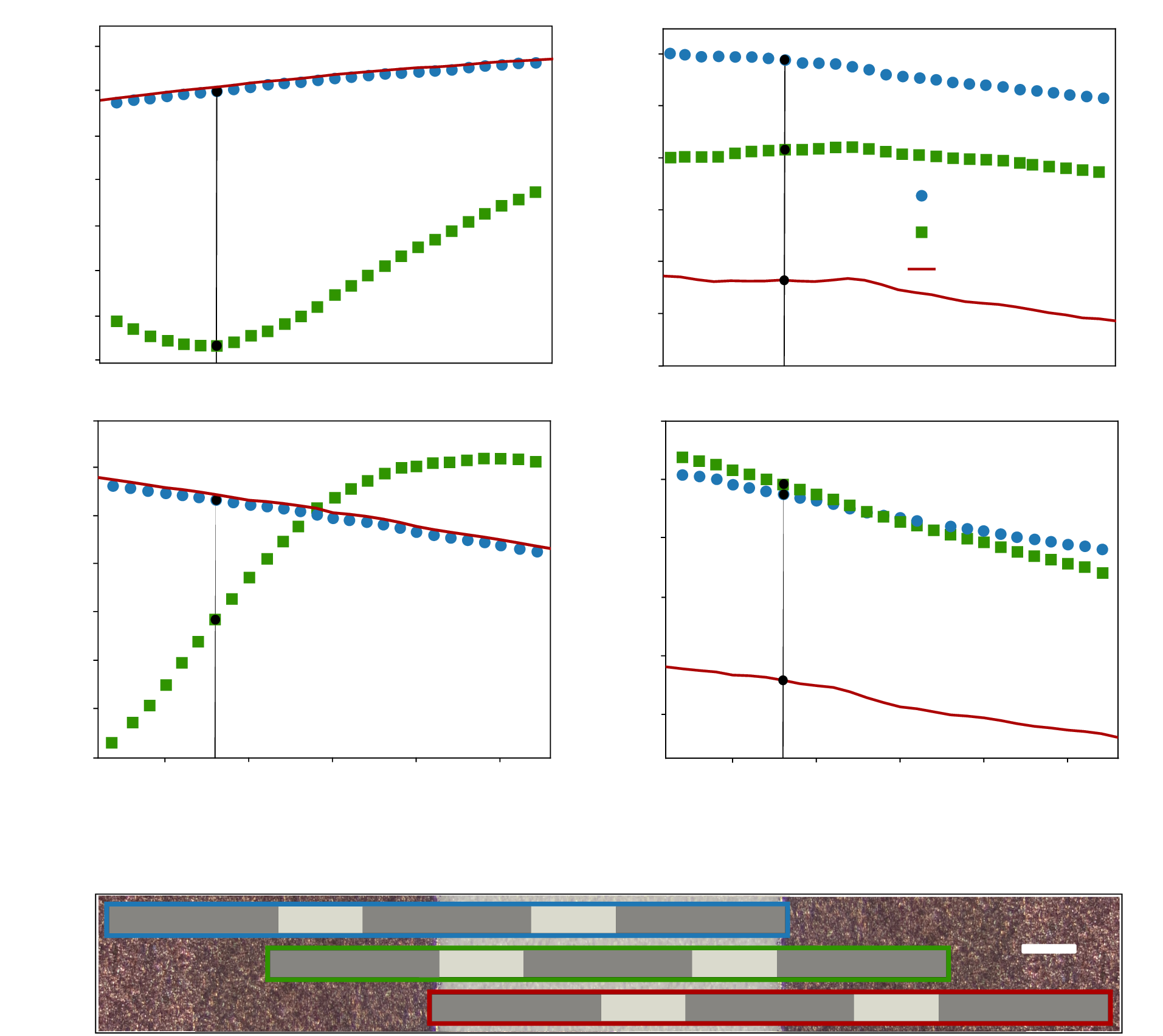
 \newline
\caption{ \textbf{Experimentally obtained reflection and transmission coefficients for three positions of the 1000 $\mu$m recess.} Amplitudes \textbf{(a,b)} and phases \textbf{(c,d)} of reflection $S_{11}$ and transmission $S_{21}$ coefficients experimentally obtained at three positions of the sensor relative to the sample \textbf{(e)} verifying an electromagnetic modeling shown in Fig.\ref{ABC_responce}. Shifted positions \textbf{A} and \textbf{C} create overlapping  $S_{11}$ amplitude and phase signals, while creating 60 deg difference in $S_{21}$ phase signal. The difference between results of electromagnetic modeling and experimentally obtained $S_{21}$ could arise from limited capabilities of our scanning system to maintain parallelism between the sensor and the sample and slight detector asymmetry due to imperfections of it's elements.}
\label{Cu_scan}
\end{figure*}

The photograph of the detector made of copper plated Rogers TMM10 laminate is presented in Fig.~\ref{detector_photo}. The device consists of a strictly symmetric coplanar half-wave resonator \textbf{1}, which is formed by the central conductor and has two identical slits with two external conductors. Such a line is analogous to a coaxial cable with a screen cut along the axis. The unperturbed field of such a line (the principal mode) is symmetrical \cite{Simons2001} with respect to the central conductor and the difference of potentials of external conductors in any cross section is zero, as it should be for a coaxial cable. The resonator is excited on this principal (symmetrical) mode by a capacitive slot \textbf{2}, which is analogous to a capacitor (of sufficiently small capacitance) in the continuation of the quarter-wave coplanar line \textbf{3} connected to the excitation port \textbf{4}. The working end of the resonator is open and the scattering field of this end is the sensitivity area \textbf{5}. The appearance of an object located near the sensitivity area changes the resonant frequency of the coplanar resonator, and the asymmetric position of the object relative to the open end of such structure breaks the symmetry of the resonator \textbf{1} field and leads to emergence of a potential difference in all cross sections of the resonator. Such a potential difference corresponds to the slot-line mode \cite{Ponchak2005}, in which the presence of the central conductor does not play an essential role. Both modes, symmetric and asymmetric, can coexist in the same line, if no means are taken to suppress one of them. As a rule, slot mode is suppressed by installing a bridge, which equalizes the potential difference between external conductors \cite{Koster1989, Kwon2001}, exactly as it happens in coaxial with a cylindrical screen. The potential difference is naturally dampened by the current arising in the bridge \textbf{6} due to the vanishingly small resistance of such a structure.

If the appearance of the object occurs symmetrically with respect to all three conductors of the coplanar waveguide, the object certainly changes the boundary condition of the resonator, but it does not break the open-end field symmetry of the resonator \textbf{1}. If such symmetry is not broken, the frequency change of the half-wave resonator occurs without the appearance of a slot mode, and the object is registered as a change in the reflected signal $S_{11}$ at the excitation port \textbf{4}.

Symmetrical reflection cannot create a potential difference of the grounds, and there is no current in the bridge (coefficient $S_{21}$ is tends to zero). This reflection-based method of data collection about a scanned sample is used by conventional TLR NSMM systems.

The slot mode characterizes the imbalance of the field in the two slits of the coplanar line \cite{Ponchak2005}. This imbalance occurs due to unequal boundary conditions for each of the slits in the presence of an object if the object is located asymmetrically relative to the two slits. The asymmetric boundary condition (asymmetric oncoming of the object) causes the appearance of the slot line mode. So, the task of the device is to intercept the current in the bridge \textbf{6} and channel this energy into the recorder through data collection port \textbf{7}, connected to a second port of a vector network analyzer.

The structure of the sensor makes it possible to receive both signals ($S_{11}$ and $S_{21}$) simultaneously at the scanning procedure, so the main goal of the study is to cooperate the coplanar waveguide ground potentials imbalance signal $S_{21}$ with the reflected from the resonator $S_{11}$ signal in order to improve TLR-based NSMM technique.

\section{Electromagnetic modeling}

 In order to built the electromagnetic model of the system AWR Design Environment was used. The geometrical parameters of the device elements were selected by means of built-in tool TXLine. This tool was utilized to transform the electrical characteristics of planar waveguides into physical ones and vice versa. We primary specified the material parameters of the substrate on which the structure was placed (dielectric permittivity, dissipation factor, dielectric thickness and metal thickness). Similar to the experimental sample based on a Rogers TMM10 laminate, the parameters were taken as follows: dielectric permittivity of 9.8, dielectric thickness of 500~$\mu$m, and copper thickness of 17~$\mu$m.

In addition to material parameters, the structure-ports impedances matching, the size of physical connectors (SMP 40) and the selected operating frequency of the sensor (7 GHz) were considered in model development.

We described above the structure of the coplanar resonator underlying the device with the slot mode pickup bridge represented as a thin metal wire between the grounds of the coplanar waveguide. Such an idealized approach is hard to be implemented directly in the final device, therefore, the unbalance current is directed into a slot line, which effect to the resonator is assumed the same as it were a recorder-loaded bridging wire. To pick up the unbalance signal to a recorder, a slot-to-microstrip transition is designed using both sides of the PCB. The slot line had galvanic contact with the coplanar grounds for capturing the potential difference. The resulting slot mode induced a current in the microstrip line, connected to the signal collection port.

After constructing the model, a metal surface defect was selected as a sample to simulate the performance of the system (Fig.~\ref{ABC_responce}). The dimensions of a square recess in a copper plate were 500x250 $\mu$m and the distance from the open end of the half-wave coplanar resonator to the object was 100 $\mu$m. Three scanned positions of the recess are shown in Figure~\ref{ABC_responce}, one of which \textbf{B} is symmetrical with respect to the central conductor of the coplanar, and the other two (\textbf{A} and \textbf{C}) are shifted relative to the central position by 100~$\mu$m to the right and left respectively.

As a result of numerical simulation, the amplitude and phase of $S_{11}$ and $S_{21}$ parameters for each of the three recess positions were obtained, as shown in Fig.~\ref{ABC_responce}.

At the shifted positions of the recess (\textbf{A} and \textbf{C}) there is an overlap of the $S_{21}$ and $S_{11}$ parameter amplitudes, since the structure of the sensitivity area is symmetrical relative to the central conductor and the imbalance of the ground potentials is the same, as well as the frequency shift of the half-wave coplanar resonator. In the case of the position \textbf{B} the imbalance of the ground potentials is much smaller (-42 dB) compared to positions \textbf{A} and \textbf{C} (-32 dB). Thus, taking the system out of equilibrium, we obtain a response of the order of 10 dB by the amplitude of transmittance parameter $S_{21}$ at -42 dBm level of signal (Fig.~\ref{ABC_responce}(b)). At the same time, the response for the reflection-based method is 7 dB at -15 dBm (Fig.~\ref{ABC_responce}(a)). So the numerical calculation demonstrates the ability to distinguish shifted (\textbf{A},\textbf{C}) and central (\textbf{B}) positions by amplitudes of both methods.

In addition we observed the possibility to distinguish two shifted positions by the phase of the $S_{21}$ parameter (Fig.~\ref{ABC_responce}(d)). The $S_{21}$-phase difference between \textbf{A} and \textbf{C} positions of recess is approximately $\pi$, which makes sense since in the case of position \textbf{A} the coplanar grounds equilibrating current flows in the opposite direction compared to position \textbf{C}. At the same time, the $S_{11}$ phases of the shifted positions overlap each other (Fig.~\ref{ABC_responce}(c)), so we cannot distinguish the positions \textbf{A} and \textbf{C} neither by the amplitude, nor the phase of the $S_{11}$ parameter.

\section{Experimental scanning}

The XYZ-scanning system was used in order to perform scanning and positioning of the sample. The scanner was controlled by an XYZ-Driver connected to a PC. Amplitude and phase of the $S_{11}$ and $S_{21}$ parameters were collected using N524A PNA-X Network Analyzer.  The measurement setup (coaxial interconnections) was calibrated with the N4690 Electronic Calibrator (ECal).

We located the sensor on the X-stage to control the distance to the sample under the scan and YZ-stage was used to 2D-movement of the sample. The XYZ-scanning system allowed to control movements at the level of 1 $\mu$m in all three axes.

A copper plated Rogers TMM10 laminate with a 1000 $\mu$m etched slot was chosen in order to verify the results of electromagnetic modeling. Positions \textbf{A} and \textbf{C}, as shown in Fig.~\ref{Cu_scan}(e), correspond to shifted location of the sensor, while position \textbf{B} corresponds to symmetrical position at Fig.~\ref{ABC_responce}(e).

The amplitudes and phases of reflection $S_{11}$ and transmission $S_{21}$ coefficients collected at three positions of the detector relative to the sample are shown in Fig.~\ref{Cu_scan}(a,b) and Fig.~\ref{Cu_scan}(c,d) correspondingly. The amplitude (Fig.~\ref{Cu_scan}(a)) and phase (Fig.~\ref{Cu_scan}(c)) overlap at the shifted positions proves the inability of S11 approach to distinguish symmetrical inhomogenities relative to the central conductor of the coplanar waveguide. The $S_{21}$ signal allows to resolve the shifted positions by the phase of the parameter (Fig.~\ref{Cu_scan}(d)). We observed in experiment less than $\pi$ phase difference and no amplitude overlap for positions \textbf{A} and \textbf{C}. We believe that the reason for this discrepancy between the experiment and electromagnetic modeling is the imperfection of the scanning system, namely the lack of precision control of the parallelism of the sensor relative to the sample and slight detector asymmetry due to imperfections of it's fabrication and mounting. This leads to an unequal imbalance of the slits in two shifted positions of the sensor and a corresponding difference in $S_{21}$ amplitudes in these positions (Fig.~\ref{Cu_scan}(b)).

\begin{figure}[tbhp]
\centering
\normalsize
 \def\svgwidth{0.48\textwidth}  
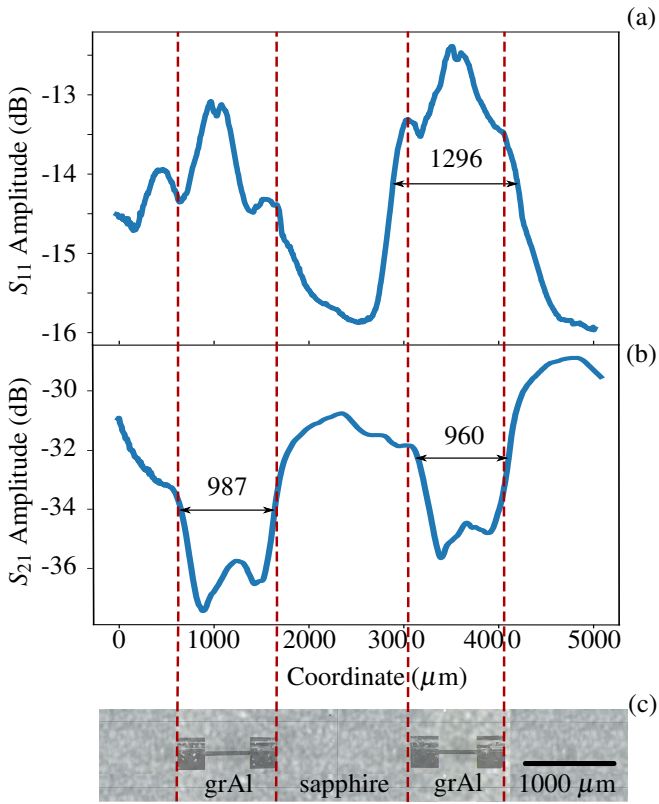
 \newline
\caption{\textbf{Line scan of granular aluminum on sapphire.} Amplitude of reflection \textbf{(a)} and transmission \textbf{(b)} coefficients for line scan of the 20 nm thick and 1000 $\mu$m wide grAl film with resistivity of 60 m$\Omega\cdot$sm sample \textbf{(c)}. The width of the grAl structures was estimated as full width of the peak (dip) at it's half-height. This levels are shown by black arrows with corresponding width on top. It's impossible to estimate the first grAl structure width by $S_{11}$ amplitude.}
\label{grAl_scan}
\end{figure}

For scanning a thin-film structure, a 20 nm thick and 1000 $\mu$m long granular aluminum (grAl) \cite{Cohen1968, Deutscher1973, Dynes1981} film with a sheet resistance of 30 kOhm on sapphire substrate was chosen (Fig.~\ref{grAl_scan}(c)). The line scan with step size of 6 $\mu$m was performed at a distance of 30 $\mu$m, and the resonant frequency of the sapphire-calibrated sensor was fixed at 6.93 GHz.

The $S_{11}$ and $S_{21}$ signal amplitudes of grAl on sapphire line scan are presented in Fig.~\ref{grAl_scan} (a) and (b) correspondingly. The widths of thin-film structures was estimated by finding the full width of a peak or a dip at half-maximum of the signal. In order to find the width of the structures from the $S_{21}$ amplitude's data, we took the extreme right maximum relative to the dip and the minimum signal value in the dip itself, with a further search for the nearest data points based on the half-differences of these two signal levels. Half-height full-width line scan data processing results in a 987 $\mu$m width of the first grAl structure and 960 $\mu$m width of the second grAl structure by $S_{21}$ amplitude. 
The calculations of the width based on $S_{11}$ amplitude's data consisted of finding the maximum amplitude of the peak and the extreme left minimum relative to the peak itself with the same procedure of half-differences signal. It was hard to distinguish the width of the first grAl structure by $S_{11}$ amplitude, therefore  Fig.~\ref{grAl_scan} (a) includes only 1296 $\mu$m width of the second structure. Best $S_{21}$  processing result have a good agreement between physical (1000 $\mu$m) and measured (987 $\mu$m) lengths of the granular aluminum thin film with difference of only 1.3 percent, while $S_{11}$ approach results in almost 30 percent difference.

\section{Conclusion}

In this paper, a NSMM platform demonstrator based on a conventional TLR-based technique with an additional source of useful signal is developed and the properties of the detector are verified using electromagnetic modeling as well as experimental scanning of the samples.

Through the electromagnetic modeling, it is concluded that both transmission-based ($S_{21}$) and reflection-based ($S_{11}$) measurements are able to distinguish central and 100 $\mu$m - shifted positions of the copper plate inhomogeneity. Moreover, transmission-based approach is able to distinguish symmetrically shifted positions of the recess by the phase of $S_{21}$ parameter since the coplanar grounds equilibrating current flows in different directions despite the symmetry of the sensing area.
In order to confirm the results of electromagnetic modeling, a Rogers TMM10 laminate with a 1000 $\mu$m etched slot was chosen.
It's experimental three-position scanning proves the possibility to distinguish shifted positions of the inhomogenity by the phase of $S_{21}$ coefficient, but the lack of control over the parallelism of the sensor  relative to the sample surface and it's imperfection lead to a decrease in the phase response and the appearance of an amplitude  response between two shifted positions.

$S_{21}$ approach is able to resolve 30 kOhm grAl structure on sapphire substrate and measure it's length with accuracy up to 98.7 \%, which is significantly higher then accuracy of $S_{11}$ approach.

Further research will focus on the improvement of the scanning system in order to maintain parallelism between the surfaces of the sensor and the sample during the scanning process, as well as on amplification of $S_{21}$ signal level and the spatial resolution of the technique. In fact, the spatial resolution of the NSMM is typically given by the sensor aperture: the smaller the sensor aperture, the higher the spatial resolution \cite{KimLee2003, Imtiaz2006}. Therefore we shall localize the sensitivity area of the sensor in order to increase the resolution and reach advanced results \cite{Anlage2001,Imtiaz2003}.

\section*{Acknowledgment}
This work was performed with the financial support of the Russian Science Foundation, Project No. 21-72-30026 (https://rscf.ru/en/project/21-72-30026/).

\section*{Data Availability}
The data that support the findings of this study are available from the corresponding author upon reasonable request.

\end{document}

%% file: fig1_tex.eps_tex
\begingroup%
  \makeatletter%
  \providecommand\color[2][]{%
    \errmessage{(Inkscape) Color is used for the text in Inkscape, but the package 'color.sty' is not loaded}%
    \renewcommand\color[2][]{}%
  }%
  \providecommand\transparent[1]{%
    \errmessage{(Inkscape) Transparency is used (non-zero) for the text in Inkscape, but the package 'transparent.sty' is not loaded}%
    \renewcommand\transparent[1]{}%
  }%
  \providecommand\rotatebox[2]{#2}%
  \newcommand*\fsize{\dimexpr\f@size pt\relax}%
  \newcommand*\lineheight[1]{\fontsize{\fsize}{#1\fsize}\selectfont}%
  \ifx\svgwidth\undefined%
    \setlength{\unitlength}{218.34911917bp}%
    \ifx\svgscale\undefined%
      \relax%
    \else%
      \setlength{\unitlength}{\unitlength * \real{\svgscale}}%
    \fi%
  \else%
    \setlength{\unitlength}{\svgwidth}%
  \fi%
  \global\let\svgwidth\undefined%
  \global\let\svgscale\undefined%
  \makeatother%
  \begin{picture}(1,1.52092507)%
    \lineheight{1}%
    \setlength\tabcolsep{0pt}%
    \put(0,0){\includegraphics[width=\unitlength]{fig1_tex.eps}}%
    \put(0.91861667,0.82583604){\color[rgb]{0,0,0}\makebox(0,0)[lt]{\lineheight{1.25}\smash{\begin{tabular}[t]{l}(c)\end{tabular}}}}%
    \put(0.94893866,0.56520628){\color[rgb]{0,0,0}\rotatebox{90}{\makebox(0,0)[lt]{\lineheight{1.25}\smash{\begin{tabular}[t]{l}VNA\end{tabular}}}}}%
    \put(0.94953645,0.69550974){\color[rgb]{0,0,0}\rotatebox{90}{\makebox(0,0)[lt]{\lineheight{1.25}\smash{\begin{tabular}[t]{l}1\end{tabular}}}}}%
    \put(0.94969555,0.50857411){\color[rgb]{0,0,0}\rotatebox{90}{\makebox(0,0)[lt]{\lineheight{1.25}\smash{\begin{tabular}[t]{l}2\end{tabular}}}}}%
    \put(0.2103093,0.04705557){\color[rgb]{0,0,0}\makebox(0,0)[lt]{\lineheight{1.25}\smash{\begin{tabular}[t]{l}Sample\end{tabular}}}}%
    \put(0.29251419,0.30597974){\color[rgb]{0,0,0}\rotatebox{90}{\makebox(0,0)[lt]{\lineheight{1.25}\smash{\begin{tabular}[t]{l}$\lambda/2$\end{tabular}}}}}%
    \put(0.28630436,0.59294326){\color[rgb]{0,0,0}\rotatebox{90}{\makebox(0,0)[lt]{\lineheight{1.25}\smash{\begin{tabular}[t]{l}$\lambda/4$\end{tabular}}}}}%
    \put(0.91867389,1.47151506){\color[rgb]{0,0,0}\makebox(0,0)[lt]{\lineheight{1.25}\smash{\begin{tabular}[t]{l}(b)\end{tabular}}}}%
    \put(0.68925579,1.47166868){\color[rgb]{0,0,0}\makebox(0,0)[lt]{\lineheight{1.25}\smash{\begin{tabular}[t]{l}6\end{tabular}}}}%
    \put(0.41552868,1.47100847){\color[rgb]{0,0,0}\makebox(0,0)[lt]{\lineheight{1.25}\smash{\begin{tabular}[t]{l}(a)\end{tabular}}}}%
    \put(0.05889464,1.45163745){\color[rgb]{0,0,0}\makebox(0,0)[lt]{\lineheight{1.25}\smash{\begin{tabular}[t]{l}7\end{tabular}}}}%
    \put(0.00885557,1.19711102){\color[rgb]{0,0,0}\makebox(0,0)[lt]{\lineheight{1.25}\smash{\begin{tabular}[t]{l}1\end{tabular}}}}%
    \put(0.36141384,1.34479644){\color[rgb]{0,0,0}\makebox(0,0)[lt]{\lineheight{1.25}\smash{\begin{tabular}[t]{l}6\end{tabular}}}}%
    \put(0.54028818,1.27410631){\color[rgb]{0,0,0}\makebox(0,0)[lt]{\lineheight{1.25}\smash{\begin{tabular}[t]{l}5\end{tabular}}}}%
    \put(0.93021969,1.20444222){\color[rgb]{0,0,0}\makebox(0,0)[lt]{\lineheight{1.25}\smash{\begin{tabular}[t]{l}4\end{tabular}}}}%
    \put(0.91062659,1.28918836){\color[rgb]{0,0,0}\makebox(0,0)[lt]{\lineheight{1.25}\smash{\begin{tabular}[t]{l}3\end{tabular}}}}%
    \put(0.88197343,1.3608916){\color[rgb]{0,0,0}\makebox(0,0)[lt]{\lineheight{1.25}\smash{\begin{tabular}[t]{l}2\end{tabular}}}}%
  \end{picture}%
\endgroup%

%% file: fig3_tex.eps_tex
\begingroup%
  \makeatletter%
  \providecommand\color[2][]{%
    \errmessage{(Inkscape) Color is used for the text in Inkscape, but the package 'color.sty' is not loaded}%
    \renewcommand\color[2][]{}%
  }%
  \providecommand\transparent[1]{%
    \errmessage{(Inkscape) Transparency is used (non-zero) for the text in Inkscape, but the package 'transparent.sty' is not loaded}%
    \renewcommand\transparent[1]{}%
  }%
  \providecommand\rotatebox[2]{#2}%
  \newcommand*\fsize{\dimexpr\f@size pt\relax}%
  \newcommand*\lineheight[1]{\fontsize{\fsize}{#1\fsize}\selectfont}%
  \ifx\svgwidth\undefined%
    \setlength{\unitlength}{914.30489127bp}%
    \ifx\svgscale\undefined%
      \relax%
    \else%
      \setlength{\unitlength}{\unitlength * \real{\svgscale}}%
    \fi%
  \else%
    \setlength{\unitlength}{\svgwidth}%
  \fi%
  \global\let\svgwidth\undefined%
  \global\let\svgscale\undefined%
  \makeatother%
  \begin{picture}(1,0.92063668)%
    \lineheight{1}%
    \setlength\tabcolsep{0pt}%
    \put(0,0){\includegraphics[width=\unitlength]{fig3_tex.eps}}%
    \put(0.27387652,0.34225419){\color[rgb]{0,0,0}\makebox(0,0)[t]{\lineheight{1.25}\smash{\begin{tabular}[t]{c}-141 deg\end{tabular}}}}%
    \put(0.24889816,0.84802546){\color[rgb]{0,0,0}\makebox(0,0)[t]{\lineheight{1.25}\smash{\begin{tabular}[t]{c}-8 dB\end{tabular}}}}%
    \put(0.26515719,0.67632131){\color[rgb]{0,0,0}\makebox(0,0)[t]{\lineheight{1.25}\smash{\begin{tabular}[t]{c}-15 dB\end{tabular}}}}%
    \put(0.49876489,0.92654269){\color[rgb]{0,0,0}\makebox(0,0)[t]{\lineheight{1.25}\smash{\begin{tabular}[t]{c}(a)\end{tabular}}}}%
    \put(0.49708299,0.58620211){\color[rgb]{0,0,0}\makebox(0,0)[t]{\lineheight{1.25}\smash{\begin{tabular}[t]{c}(c)\end{tabular}}}}%
    \put(0.76617664,0.54305987){\color[rgb]{0,0,0}\makebox(0,0)[t]{\lineheight{1.25}\smash{\begin{tabular}[t]{c}109 deg \end{tabular}}}}%
    \put(0.76641175,0.31634573){\color[rgb]{0,0,0}\makebox(0,0)[t]{\lineheight{1.25}\smash{\begin{tabular}[t]{c}-148 deg\end{tabular}}}}%
    \put(0.27029683,0.480622){\color[rgb]{0,0,0}\makebox(0,0)[t]{\lineheight{1.25}\smash{\begin{tabular}[t]{c}-58 deg\end{tabular}}}}%
    \put(0.76584836,0.38020507){\color[rgb]{0,0,0}\makebox(0,0)[t]{\lineheight{1.25}\smash{\begin{tabular}[t]{c}-102 deg\end{tabular}}}}%
    \put(0.99069402,0.92552338){\color[rgb]{0,0,0}\makebox(0,0)[t]{\lineheight{1.25}\smash{\begin{tabular}[t]{c}(b)\end{tabular}}}}%
    \put(0.75200741,0.88043742){\color[rgb]{0,0,0}\makebox(0,0)[t]{\lineheight{1.25}\smash{\begin{tabular}[t]{c}-32 dB\end{tabular}}}}%
    \put(0.75542081,0.64586796){\color[rgb]{0,0,0}\makebox(0,0)[t]{\lineheight{1.25}\smash{\begin{tabular}[t]{c}-42 dB\end{tabular}}}}%
    \put(0.99112899,0.59229248){\color[rgb]{0,0,0}\makebox(0,0)[t]{\lineheight{1.25}\smash{\begin{tabular}[t]{c}(d)\end{tabular}}}}%
    \put(0.99108796,0.18326305){\color[rgb]{0,0,0}\makebox(0,0)[t]{\lineheight{1.25}\smash{\begin{tabular}[t]{c}(e)\end{tabular}}}}%
    \put(0.1015437,0.04915028){\color[rgb]{0,0,0}\makebox(0,0)[t]{\lineheight{1.25}\smash{\begin{tabular}[t]{c}A\end{tabular}}}}%
    \put(0.40279777,0.048754){\color[rgb]{0,0,0}\makebox(0,0)[t]{\lineheight{1.25}\smash{\begin{tabular}[t]{c}B\end{tabular}}}}%
    \put(0.70409901,0.0489965){\color[rgb]{0,0,0}\makebox(0,0)[t]{\lineheight{1.25}\smash{\begin{tabular}[t]{c}C\end{tabular}}}}%
    \put(0.10123183,0.23484277){\color[rgb]{0,0,0}\makebox(0,0)[t]{\lineheight{1.25}\smash{\begin{tabular}[t]{c}6.660\end{tabular}}}}%
    \put(0.05081468,0.28088086){\color[rgb]{0,0,0}\makebox(0,0)[t]{\lineheight{1.25}\smash{\begin{tabular}[t]{c}-160\\\end{tabular}}}}%
    \put(0.54376655,0.29291814){\color[rgb]{0,0,0}\makebox(0,0)[t]{\lineheight{1.25}\smash{\begin{tabular}[t]{c}-150\\\end{tabular}}}}%
    \put(0.05081468,0.31660305){\color[rgb]{0,0,0}\makebox(0,0)[t]{\lineheight{1.25}\smash{\begin{tabular}[t]{c}-140\\\end{tabular}}}}%
    \put(0.05081468,0.3523252){\color[rgb]{0,0,0}\makebox(0,0)[t]{\lineheight{1.25}\smash{\begin{tabular}[t]{c}-120\\\end{tabular}}}}%
    \put(0.05081468,0.38794484){\color[rgb]{0,0,0}\makebox(0,0)[t]{\lineheight{1.25}\smash{\begin{tabular}[t]{c}-100\\\end{tabular}}}}%
    \put(0.05509565,0.42376954){\color[rgb]{0,0,0}\makebox(0,0)[t]{\lineheight{1.25}\smash{\begin{tabular}[t]{c}-80\\\end{tabular}}}}%
    \put(0.05509565,0.45949171){\color[rgb]{0,0,0}\makebox(0,0)[t]{\lineheight{1.25}\smash{\begin{tabular}[t]{c}-60\\\end{tabular}}}}%
    \put(0.05509565,0.49521387){\color[rgb]{0,0,0}\makebox(0,0)[t]{\lineheight{1.25}\smash{\begin{tabular}[t]{c}-40\\\end{tabular}}}}%
    \put(0.05509565,0.53093603){\color[rgb]{0,0,0}\makebox(0,0)[t]{\lineheight{1.25}\smash{\begin{tabular}[t]{c}-20\\\end{tabular}}}}%
    \put(0.17515879,0.23484277){\color[rgb]{0,0,0}\makebox(0,0)[t]{\lineheight{1.25}\smash{\begin{tabular}[t]{c}6.665\end{tabular}}}}%
    \put(0.24918326,0.23484277){\color[rgb]{0,0,0}\makebox(0,0)[t]{\lineheight{1.25}\smash{\begin{tabular}[t]{c}6.670\end{tabular}}}}%
    \put(0.32311025,0.23484277){\color[rgb]{0,0,0}\makebox(0,0)[t]{\lineheight{1.25}\smash{\begin{tabular}[t]{c}6.675\end{tabular}}}}%
    \put(0.3971347,0.23484277){\color[rgb]{0,0,0}\makebox(0,0)[t]{\lineheight{1.25}\smash{\begin{tabular}[t]{c}6.680\end{tabular}}}}%
    \put(0.47106171,0.23484277){\color[rgb]{0,0,0}\makebox(0,0)[t]{\lineheight{1.25}\smash{\begin{tabular}[t]{c}6.685\end{tabular}}}}%
    \put(0.59104006,0.23484277){\color[rgb]{0,0,0}\makebox(0,0)[t]{\lineheight{1.25}\smash{\begin{tabular}[t]{c}6.660\end{tabular}}}}%
    \put(0.66496707,0.23484277){\color[rgb]{0,0,0}\makebox(0,0)[t]{\lineheight{1.25}\smash{\begin{tabular}[t]{c}6.665\end{tabular}}}}%
    \put(0.73899147,0.23484277){\color[rgb]{0,0,0}\makebox(0,0)[t]{\lineheight{1.25}\smash{\begin{tabular}[t]{c}6.670\end{tabular}}}}%
    \put(0.81291853,0.23484277){\color[rgb]{0,0,0}\makebox(0,0)[t]{\lineheight{1.25}\smash{\begin{tabular}[t]{c}6.675\end{tabular}}}}%
    \put(0.88694298,0.23484277){\color[rgb]{0,0,0}\makebox(0,0)[t]{\lineheight{1.25}\smash{\begin{tabular}[t]{c}6.680\end{tabular}}}}%
    \put(0.96086994,0.23484277){\color[rgb]{0,0,0}\makebox(0,0)[t]{\lineheight{1.25}\smash{\begin{tabular}[t]{c}6.685\end{tabular}}}}%
    \put(0.27891674,0.1971589){\color[rgb]{0,0,0}\makebox(0,0)[t]{\lineheight{1.25}\smash{\begin{tabular}[t]{c}Frequency (GHz)\end{tabular}}}}%
    \put(0.76872498,0.19751075){\color[rgb]{0,0,0}\makebox(0,0)[t]{\lineheight{1.25}\smash{\begin{tabular}[t]{c}Frequency (GHz)\end{tabular}}}}%
    \put(0.01240436,0.41495084){\color[rgb]{0,0,0}\rotatebox{90}{\makebox(0,0)[t]{\lineheight{1.25}\smash{\begin{tabular}[t]{c}$S_{11}$ Phase (deg)\end{tabular}}}}}%
    \put(0.01240436,0.75398212){\color[rgb]{0,0,0}\rotatebox{90}{\makebox(0,0)[t]{\lineheight{1.25}\smash{\begin{tabular}[t]{c}$S_{11}$ Amplitude (dB)\end{tabular}}}}}%
    \put(0.05559731,0.65751028){\color[rgb]{0,0,0}\makebox(0,0)[t]{\lineheight{1.25}\smash{\begin{tabular}[t]{c}-14\\\end{tabular}}}}%
    \put(0.05554908,0.60635839){\color[rgb]{0,0,0}\makebox(0,0)[t]{\lineheight{1.25}\smash{\begin{tabular}[t]{c}-16\\\end{tabular}}}}%
    \put(0.05554908,0.70854135){\color[rgb]{0,0,0}\makebox(0,0)[t]{\lineheight{1.25}\smash{\begin{tabular}[t]{c}-12\\\end{tabular}}}}%
    \put(0.05567904,0.76066078){\color[rgb]{0,0,0}\makebox(0,0)[t]{\lineheight{1.25}\smash{\begin{tabular}[t]{c}-10\\\end{tabular}}}}%
    \put(0.05918609,0.81089282){\color[rgb]{0,0,0}\makebox(0,0)[t]{\lineheight{1.25}\smash{\begin{tabular}[t]{c}-8\\\end{tabular}}}}%
    \put(0.05916173,0.86341025){\color[rgb]{0,0,0}\makebox(0,0)[t]{\lineheight{1.25}\smash{\begin{tabular}[t]{c}-6\\\end{tabular}}}}%
    \put(0.54376693,0.34027255){\color[rgb]{0,0,0}\makebox(0,0)[t]{\lineheight{1.25}\smash{\begin{tabular}[t]{c}-100\\\end{tabular}}}}%
    \put(0.54376693,0.38731385){\color[rgb]{0,0,0}\makebox(0,0)[t]{\lineheight{1.25}\smash{\begin{tabular}[t]{c}-50\\\end{tabular}}}}%
    \put(0.54376693,0.43466516){\color[rgb]{0,0,0}\makebox(0,0)[t]{\lineheight{1.25}\smash{\begin{tabular}[t]{c}0\\\end{tabular}}}}%
    \put(0.54376693,0.48170955){\color[rgb]{0,0,0}\makebox(0,0)[t]{\lineheight{1.25}\smash{\begin{tabular}[t]{c}50\\\end{tabular}}}}%
    \put(0.54376693,0.52906087){\color[rgb]{0,0,0}\makebox(0,0)[t]{\lineheight{1.25}\smash{\begin{tabular}[t]{c}100\\\end{tabular}}}}%
    \put(0.50969972,0.41278512){\color[rgb]{0,0,0}\rotatebox{90}{\makebox(0,0)[t]{\lineheight{1.25}\smash{\begin{tabular}[t]{c}$S_{21}$ Phase (deg)\end{tabular}}}}}%
    \put(0.50698418,0.75156415){\color[rgb]{0,0,0}\rotatebox{90}{\makebox(0,0)[t]{\lineheight{1.25}\smash{\begin{tabular}[t]{c}$S_{21}$ Amplitude (dB)\end{tabular}}}}}%
    \put(0.54052779,0.62960708){\color[rgb]{0,0,0}\makebox(0,0)[t]{\lineheight{1.25}\smash{\begin{tabular}[t]{c}-42\\\end{tabular}}}}%
    \put(0.54058563,0.67950205){\color[rgb]{0,0,0}\makebox(0,0)[t]{\lineheight{1.25}\smash{\begin{tabular}[t]{c}-40\\\\\end{tabular}}}}%
    \put(0.54062909,0.72931613){\color[rgb]{0,0,0}\makebox(0,0)[t]{\lineheight{1.25}\smash{\begin{tabular}[t]{c}-38\\\\\end{tabular}}}}%
    \put(0.54029707,0.77912681){\color[rgb]{0,0,0}\makebox(0,0)[t]{\lineheight{1.25}\smash{\begin{tabular}[t]{c}-36\\\\\end{tabular}}}}%
    \put(0.54051346,0.82893753){\color[rgb]{0,0,0}\makebox(0,0)[t]{\lineheight{1.25}\smash{\begin{tabular}[t]{c}-34\\\\\end{tabular}}}}%
    \put(0.54039456,0.87874822){\color[rgb]{0,0,0}\makebox(0,0)[t]{\lineheight{1.25}\smash{\begin{tabular}[t]{c}-32\\\\\end{tabular}}}}%
    \put(0.85143182,0.77283604){\color[rgb]{0,0,0}\makebox(0,0)[t]{\lineheight{1.25}\smash{\begin{tabular}[t]{c}A position\\\end{tabular}}}}%
    \put(0.85094192,0.74111041){\color[rgb]{0,0,0}\makebox(0,0)[t]{\lineheight{1.25}\smash{\begin{tabular}[t]{c}B position\\\end{tabular}}}}%
    \put(0.85147247,0.7091899){\color[rgb]{0,0,0}\makebox(0,0)[t]{\lineheight{1.25}\smash{\begin{tabular}[t]{c}C position\\\end{tabular}}}}%
  \end{picture}%
\endgroup%

%% file: fig2_tex.eps_tex
\begingroup%
  \makeatletter%
  \providecommand\color[2][]{%
    \errmessage{(Inkscape) Color is used for the text in Inkscape, but the package 'color.sty' is not loaded}%
    \renewcommand\color[2][]{}%
  }%
  \providecommand\transparent[1]{%
    \errmessage{(Inkscape) Transparency is used (non-zero) for the text in Inkscape, but the package 'transparent.sty' is not loaded}%
    \renewcommand\transparent[1]{}%
  }%
  \providecommand\rotatebox[2]{#2}%
  \newcommand*\fsize{\dimexpr\f@size pt\relax}%
  \newcommand*\lineheight[1]{\fontsize{\fsize}{#1\fsize}\selectfont}%
  \ifx\svgwidth\undefined%
    \setlength{\unitlength}{917.11747778bp}%
    \ifx\svgscale\undefined%
      \relax%
    \else%
      \setlength{\unitlength}{\unitlength * \real{\svgscale}}%
    \fi%
  \else%
    \setlength{\unitlength}{\svgwidth}%
  \fi%
  \global\let\svgwidth\undefined%
  \global\let\svgscale\undefined%
  \makeatother%
  \begin{picture}(1,0.89139484)%
    \lineheight{1}%
    \setlength\tabcolsep{0pt}%
    \put(0,0){\includegraphics[width=\unitlength]{fig2_tex.eps}}%
    \put(0.25156434,0.34179006){\color[rgb]{0,0,0}\makebox(0,0)[t]{\lineheight{1.25}\smash{\begin{tabular}[t]{c}-112 deg\end{tabular}}}}%
    \put(0.22761344,0.83675966){\color[rgb]{0,0,0}\makebox(0,0)[t]{\lineheight{1.25}\smash{\begin{tabular}[t]{c}-4 dB\end{tabular}}}}%
    \put(0.23413681,0.64057234){\color[rgb]{0,0,0}\makebox(0,0)[t]{\lineheight{1.25}\smash{\begin{tabular}[t]{c}-15 dB\end{tabular}}}}%
    \put(0.49883217,0.89580899){\color[rgb]{0,0,0}\makebox(0,0)[t]{\lineheight{1.25}\smash{\begin{tabular}[t]{c}(a)\end{tabular}}}}%
    \put(0.49715543,0.55651213){\color[rgb]{0,0,0}\makebox(0,0)[t]{\lineheight{1.25}\smash{\begin{tabular}[t]{c}(c)\end{tabular}}}}%
    \put(0.71872966,0.49166376){\color[rgb]{0,0,0}\makebox(0,0)[t]{\lineheight{1.25}\smash{\begin{tabular}[t]{c}81 deg \end{tabular}}}}%
    \put(0.71682354,0.31984177){\color[rgb]{0,0,0}\makebox(0,0)[t]{\lineheight{1.25}\smash{\begin{tabular}[t]{c}18 deg\end{tabular}}}}%
    \put(0.24508059,0.48056229){\color[rgb]{0,0,0}\makebox(0,0)[t]{\lineheight{1.25}\smash{\begin{tabular}[t]{c}-82 deg\end{tabular}}}}%
    \put(0.7174925,0.42695847){\color[rgb]{0,0,0}\makebox(0,0)[t]{\lineheight{1.25}\smash{\begin{tabular}[t]{c}78 deg\end{tabular}}}}%
    \put(0.98925276,0.8947928){\color[rgb]{0,0,0}\makebox(0,0)[t]{\lineheight{1.25}\smash{\begin{tabular}[t]{c}(b)\end{tabular}}}}%
    \put(0.7141738,0.84769297){\color[rgb]{0,0,0}\makebox(0,0)[t]{\lineheight{1.25}\smash{\begin{tabular}[t]{c}-26 dB\end{tabular}}}}%
    \put(0.71343229,0.6671147){\color[rgb]{0,0,0}\makebox(0,0)[t]{\lineheight{1.25}\smash{\begin{tabular}[t]{c}-34 dB\end{tabular}}}}%
    \put(0.71352734,0.78616885){\color[rgb]{0,0,0}\makebox(0,0)[t]{\lineheight{1.25}\smash{\begin{tabular}[t]{c}-30 dB\end{tabular}}}}%
    \put(0.98968649,0.56258379){\color[rgb]{0,0,0}\makebox(0,0)[t]{\lineheight{1.25}\smash{\begin{tabular}[t]{c}(d)\end{tabular}}}}%
    \put(0.98964549,0.15480879){\color[rgb]{0,0,0}\makebox(0,0)[t]{\lineheight{1.25}\smash{\begin{tabular}[t]{c}(e)\end{tabular}}}}%
    \put(0.14195951,0.20489418){\color[rgb]{0,0,0}\makebox(0,0)[t]{\lineheight{1.25}\smash{\begin{tabular}[t]{c}6.910\end{tabular}}}}%
    \put(0.05169352,0.23393163){\color[rgb]{0,0,0}\makebox(0,0)[t]{\lineheight{1.25}\smash{\begin{tabular}[t]{c}-140\\\end{tabular}}}}%
    \put(0.54390064,0.27171656){\color[rgb]{0,0,0}\makebox(0,0)[t]{\lineheight{1.25}\smash{\begin{tabular}[t]{c}0\\\end{tabular}}}}%
    \put(0.54359399,0.32192075){\color[rgb]{0,0,0}\makebox(0,0)[t]{\lineheight{1.25}\smash{\begin{tabular}[t]{c}20\\\end{tabular}}}}%
    \put(0.05169352,0.27659765){\color[rgb]{0,0,0}\makebox(0,0)[t]{\lineheight{1.25}\smash{\begin{tabular}[t]{c}-130\\\end{tabular}}}}%
    \put(0.05174462,0.31803689){\color[rgb]{0,0,0}\makebox(0,0)[t]{\lineheight{1.25}\smash{\begin{tabular}[t]{c}-120\\\end{tabular}}}}%
    \put(0.05158766,0.35990173){\color[rgb]{0,0,0}\makebox(0,0)[t]{\lineheight{1.25}\smash{\begin{tabular}[t]{c}-110\\\end{tabular}}}}%
    \put(0.05155571,0.40136703){\color[rgb]{0,0,0}\makebox(0,0)[t]{\lineheight{1.25}\smash{\begin{tabular}[t]{c}-100\\\end{tabular}}}}%
    \put(0.05586426,0.44245848){\color[rgb]{0,0,0}\makebox(0,0)[t]{\lineheight{1.25}\smash{\begin{tabular}[t]{c}-90\\\end{tabular}}}}%
    \put(0.05581979,0.48420645){\color[rgb]{0,0,0}\makebox(0,0)[t]{\lineheight{1.25}\smash{\begin{tabular}[t]{c}-80\\\end{tabular}}}}%
    \put(0.05596135,0.52405781){\color[rgb]{0,0,0}\makebox(0,0)[t]{\lineheight{1.25}\smash{\begin{tabular}[t]{c}-70\\\end{tabular}}}}%
    \put(0.21405024,0.2049025){\color[rgb]{0,0,0}\makebox(0,0)[t]{\lineheight{1.25}\smash{\begin{tabular}[t]{c}6.915\end{tabular}}}}%
    \put(0.28444712,0.20487255){\color[rgb]{0,0,0}\makebox(0,0)[t]{\lineheight{1.25}\smash{\begin{tabular}[t]{c}6.920\end{tabular}}}}%
    \put(0.35651481,0.20492626){\color[rgb]{0,0,0}\makebox(0,0)[t]{\lineheight{1.25}\smash{\begin{tabular}[t]{c}6.925\end{tabular}}}}%
    \put(0.42856727,0.20497629){\color[rgb]{0,0,0}\makebox(0,0)[t]{\lineheight{1.25}\smash{\begin{tabular}[t]{c}6.930\end{tabular}}}}%
    \put(0.2796583,0.16866201){\color[rgb]{0,0,0}\makebox(0,0)[t]{\lineheight{1.25}\smash{\begin{tabular}[t]{c}Frequency (GHz)\end{tabular}}}}%
    \put(0.76796442,0.16901279){\color[rgb]{0,0,0}\makebox(0,0)[t]{\lineheight{1.25}\smash{\begin{tabular}[t]{c}Frequency (GHz)\end{tabular}}}}%
    \put(0.01396321,0.38578605){\color[rgb]{0,0,0}\rotatebox{90}{\makebox(0,0)[t]{\lineheight{1.25}\smash{\begin{tabular}[t]{c}$S_{11}$ Phase (deg)\end{tabular}}}}}%
    \put(0.01396321,0.7237776){\color[rgb]{0,0,0}\rotatebox{90}{\makebox(0,0)[t]{\lineheight{1.25}\smash{\begin{tabular}[t]{c}$S_{11}$ Amplitude (dB)\end{tabular}}}}}%
    \put(0.05702364,0.61502449){\color[rgb]{0,0,0}\makebox(0,0)[t]{\lineheight{1.25}\smash{\begin{tabular}[t]{c}-14\\\end{tabular}}}}%
    \put(0.05697562,0.57660656){\color[rgb]{0,0,0}\makebox(0,0)[t]{\lineheight{1.25}\smash{\begin{tabular}[t]{c}-16\\\end{tabular}}}}%
    \put(0.05707783,0.6539428){\color[rgb]{0,0,0}\makebox(0,0)[t]{\lineheight{1.25}\smash{\begin{tabular}[t]{c}-12\\\end{tabular}}}}%
    \put(0.05720738,0.69261346){\color[rgb]{0,0,0}\makebox(0,0)[t]{\lineheight{1.25}\smash{\begin{tabular}[t]{c}-10\\\end{tabular}}}}%
    \put(0.06049917,0.7701893){\color[rgb]{0,0,0}\makebox(0,0)[t]{\lineheight{1.25}\smash{\begin{tabular}[t]{c}-6\\\end{tabular}}}}%
    \put(0.06081363,0.80886689){\color[rgb]{0,0,0}\makebox(0,0)[t]{\lineheight{1.25}\smash{\begin{tabular}[t]{c}-4\\\end{tabular}}}}%
    \put(0.06069927,0.73168893){\color[rgb]{0,0,0}\makebox(0,0)[t]{\lineheight{1.25}\smash{\begin{tabular}[t]{c}-8\\\end{tabular}}}}%
    \put(0.06067942,0.84769239){\color[rgb]{0,0,0}\makebox(0,0)[t]{\lineheight{1.25}\smash{\begin{tabular}[t]{c}-2\\\end{tabular}}}}%
    \put(0.54390064,0.37223874){\color[rgb]{0,0,0}\makebox(0,0)[t]{\lineheight{1.25}\smash{\begin{tabular}[t]{c}40\\\end{tabular}}}}%
    \put(0.5435173,0.52383371){\color[rgb]{0,0,0}\makebox(0,0)[t]{\lineheight{1.25}\smash{\begin{tabular}[t]{c}100\\\end{tabular}}}}%
    \put(0.54347061,0.42377581){\color[rgb]{0,0,0}\makebox(0,0)[t]{\lineheight{1.25}\smash{\begin{tabular}[t]{c}60\\\end{tabular}}}}%
    \put(0.54361771,0.47377629){\color[rgb]{0,0,0}\makebox(0,0)[t]{\lineheight{1.25}\smash{\begin{tabular}[t]{c}80\\\end{tabular}}}}%
    \put(0.50973356,0.38362694){\color[rgb]{0,0,0}\rotatebox{90}{\makebox(0,0)[t]{\lineheight{1.25}\smash{\begin{tabular}[t]{c}$S_{21}$ Phase (deg)\end{tabular}}}}}%
    \put(0.50702635,0.72136703){\color[rgb]{0,0,0}\rotatebox{90}{\makebox(0,0)[t]{\lineheight{1.25}\smash{\begin{tabular}[t]{c}$S_{21}$ Amplitude (dB)\end{tabular}}}}}%
    \put(0.54039478,0.57131755){\color[rgb]{0,0,0}\makebox(0,0)[t]{\lineheight{1.25}\smash{\begin{tabular}[t]{c}-38\\\end{tabular}}}}%
    \put(0.54014574,0.61656173){\color[rgb]{0,0,0}\makebox(0,0)[t]{\lineheight{1.25}\smash{\begin{tabular}[t]{c}-36\\\\\end{tabular}}}}%
    \put(0.54015912,0.66156938){\color[rgb]{0,0,0}\makebox(0,0)[t]{\lineheight{1.25}\smash{\begin{tabular}[t]{c}-34\\\\\end{tabular}}}}%
    \put(0.54003256,0.75058298){\color[rgb]{0,0,0}\makebox(0,0)[t]{\lineheight{1.25}\smash{\begin{tabular}[t]{c}-30\\\\\end{tabular}}}}%
    \put(0.54007911,0.70583){\color[rgb]{0,0,0}\makebox(0,0)[t]{\lineheight{1.25}\smash{\begin{tabular}[t]{c}-32\\\\\end{tabular}}}}%
    \put(0.54024829,0.79553872){\color[rgb]{0,0,0}\makebox(0,0)[t]{\lineheight{1.25}\smash{\begin{tabular}[t]{c}-28\\\\\end{tabular}}}}%
    \put(0.54033418,0.83998334){\color[rgb]{0,0,0}\makebox(0,0)[t]{\lineheight{1.25}\smash{\begin{tabular}[t]{c}-26\\\\\end{tabular}}}}%
    \put(0.85918125,0.71903451){\color[rgb]{0,0,0}\makebox(0,0)[t]{\lineheight{1.25}\smash{\begin{tabular}[t]{c}A position\\\end{tabular}}}}%
    \put(0.85869286,0.68740619){\color[rgb]{0,0,0}\makebox(0,0)[t]{\lineheight{1.25}\smash{\begin{tabular}[t]{c}B position\\\end{tabular}}}}%
    \put(0.85922178,0.65558354){\color[rgb]{0,0,0}\makebox(0,0)[t]{\lineheight{1.25}\smash{\begin{tabular}[t]{c}C position\\\end{tabular}}}}%
    \put(0.63061284,0.20459128){\color[rgb]{0,0,0}\makebox(0,0)[t]{\lineheight{1.25}\smash{\begin{tabular}[t]{c}6.910\end{tabular}}}}%
    \put(0.70270355,0.2045996){\color[rgb]{0,0,0}\makebox(0,0)[t]{\lineheight{1.25}\smash{\begin{tabular}[t]{c}6.915\end{tabular}}}}%
    \put(0.77310043,0.20456965){\color[rgb]{0,0,0}\makebox(0,0)[t]{\lineheight{1.25}\smash{\begin{tabular}[t]{c}6.920\end{tabular}}}}%
    \put(0.84516807,0.20462336){\color[rgb]{0,0,0}\makebox(0,0)[t]{\lineheight{1.25}\smash{\begin{tabular}[t]{c}6.925\end{tabular}}}}%
    \put(0.91722052,0.20467339){\color[rgb]{0,0,0}\makebox(0,0)[t]{\lineheight{1.25}\smash{\begin{tabular}[t]{c}6.930\end{tabular}}}}%
    \put(0.10694516,0.0947126){\color[rgb]{0,0,0}\makebox(0,0)[t]{\lineheight{1.25}\smash{\begin{tabular}[t]{c}A\end{tabular}}}}%
    \put(0.38421484,0.01879066){\color[rgb]{0,0,0}\makebox(0,0)[t]{\lineheight{1.25}\smash{\begin{tabular}[t]{c}C\end{tabular}}}}%
    \put(0.24845368,0.05688476){\color[rgb]{0,0,0}\makebox(0,0)[t]{\lineheight{1.25}\smash{\begin{tabular}[t]{c}B\end{tabular}}}}%
    \put(0.86111829,0.09401996){\color[rgb]{1,1,1}\makebox(0,0)[lt]{\lineheight{1.25}\smash{\begin{tabular}[t]{l}100  $\mu$m \end{tabular}}}}%
  \end{picture}%
\endgroup%

%% file: fig4_tex.eps_tex
\begingroup%
  \makeatletter%
  \providecommand\color[2][]{%
    \errmessage{(Inkscape) Color is used for the text in Inkscape, but the package 'color.sty' is not loaded}%
    \renewcommand\color[2][]{}%
  }%
  \providecommand\transparent[1]{%
    \errmessage{(Inkscape) Transparency is used (non-zero) for the text in Inkscape, but the package 'transparent.sty' is not loaded}%
    \renewcommand\transparent[1]{}%
  }%
  \providecommand\rotatebox[2]{#2}%
  \newcommand*\fsize{\dimexpr\f@size pt\relax}%
  \newcommand*\lineheight[1]{\fontsize{\fsize}{#1\fsize}\selectfont}%
  \ifx\svgwidth\undefined%
    \setlength{\unitlength}{387.58628064bp}%
    \ifx\svgscale\undefined%
      \relax%
    \else%
      \setlength{\unitlength}{\unitlength * \real{\svgscale}}%
    \fi%
  \else%
    \setlength{\unitlength}{\svgwidth}%
  \fi%
  \global\let\svgwidth\undefined%
  \global\let\svgscale\undefined%
  \makeatother%
  \begin{picture}(1,1.19816721)%
    \lineheight{1}%
    \setlength\tabcolsep{0pt}%
    \put(0,0){\includegraphics[width=\unitlength]{fig4_tex.eps}}%
    \put(0.97797407,0.68473577){\color[rgb]{0,0,0}\makebox(0,0)[t]{\lineheight{1.25}\smash{\begin{tabular}[t]{c}(b)\end{tabular}}}}%
    \put(0.98049184,0.14251232){\color[rgb]{0,0,0}\makebox(0,0)[t]{\lineheight{1.25}\smash{\begin{tabular}[t]{c}(c)\end{tabular}}}}%
    \put(0.02852841,0.49907782){\color[rgb]{0,0,0}\rotatebox{90}{\makebox(0,0)[t]{\lineheight{1.25}\smash{\begin{tabular}[t]{c}$S_{21}$ Amplitude (dB)\end{tabular}}}}}%
    \put(0.07656733,0.44740315){\color[rgb]{0,0,0}\makebox(0,0)[t]{\lineheight{1.25}\smash{\begin{tabular}[t]{c}-34\end{tabular}}}}%
    \put(0.07691665,0.53843046){\color[rgb]{0,0,0}\makebox(0,0)[t]{\lineheight{1.25}\smash{\begin{tabular}[t]{c}-32\\\end{tabular}}}}%
    \put(0.07869166,0.63068439){\color[rgb]{0,0,0}\makebox(0,0)[t]{\lineheight{1.25}\smash{\begin{tabular}[t]{c}-30\\\end{tabular}}}}%
    \put(0.07760076,0.35576266){\color[rgb]{0,0,0}\makebox(0,0)[t]{\lineheight{1.25}\smash{\begin{tabular}[t]{c}-36\\\end{tabular}}}}%
    \put(0.1710309,0.24142104){\color[rgb]{0,0,0}\makebox(0,0)[t]{\lineheight{1.25}\smash{\begin{tabular}[t]{c}0\end{tabular}}}}%
    \put(0.31635369,0.24142104){\color[rgb]{0,0,0}\makebox(0,0)[t]{\lineheight{1.25}\smash{\begin{tabular}[t]{c}1000\end{tabular}}}}%
    \put(0.46479381,0.24142104){\color[rgb]{0,0,0}\makebox(0,0)[t]{\lineheight{1.25}\smash{\begin{tabular}[t]{c}2000\end{tabular}}}}%
    \put(0.58569359,0.24142104){\color[rgb]{0,0,0}\makebox(0,0)[t]{\lineheight{1.25}\smash{\begin{tabular}[t]{c}300\end{tabular}}}}%
    \put(0.34023455,0.0200995){\color[rgb]{0,0,0}\makebox(0,0)[t]{\lineheight{1.25}\smash{\begin{tabular}[t]{c}grAl\end{tabular}}}}%
    \put(0.69492935,0.02062707){\color[rgb]{0,0,0}\makebox(0,0)[t]{\lineheight{1.25}\smash{\begin{tabular}[t]{c}grAl\end{tabular}}}}%
    \put(0.51909908,0.01981085){\color[rgb]{0,0,0}\makebox(0,0)[t]{\lineheight{1.25}\smash{\begin{tabular}[t]{c}sapphire\end{tabular}}}}%
    \put(0.63342765,0.24142104){\color[rgb]{0,0,0}\makebox(0,0)[t]{\lineheight{1.25}\smash{\begin{tabular}[t]{c}0\end{tabular}}}}%
    \put(0.73438568,0.24142104){\color[rgb]{0,0,0}\makebox(0,0)[t]{\lineheight{1.25}\smash{\begin{tabular}[t]{c}400\end{tabular}}}}%
    \put(0.78211974,0.24142104){\color[rgb]{0,0,0}\makebox(0,0)[t]{\lineheight{1.25}\smash{\begin{tabular}[t]{c}0\end{tabular}}}}%
    \put(0.90502225,0.24142104){\color[rgb]{0,0,0}\makebox(0,0)[t]{\lineheight{1.25}\smash{\begin{tabular}[t]{c}5000\end{tabular}}}}%
    \put(0.33851521,0.47902982){\color[rgb]{0,0,0}\makebox(0,0)[t]{\lineheight{1.25}\smash{\begin{tabular}[t]{c}987\end{tabular}}}}%
    \put(0.70420782,0.55725349){\color[rgb]{0,0,0}\makebox(0,0)[t]{\lineheight{1.25}\smash{\begin{tabular}[t]{c}960\end{tabular}}}}%
    \put(0.07825653,0.71684725){\color[rgb]{0,0,0}\makebox(0,0)[t]{\lineheight{1.25}\smash{\begin{tabular}[t]{c}-16\end{tabular}}}}%
    \put(0.0782565,0.83960472){\color[rgb]{0,0,0}\makebox(0,0)[t]{\lineheight{1.25}\smash{\begin{tabular}[t]{c}-15\end{tabular}}}}%
    \put(0.0782565,0.96236756){\color[rgb]{0,0,0}\makebox(0,0)[t]{\lineheight{1.25}\smash{\begin{tabular}[t]{c}-14\end{tabular}}}}%
    \put(0.0782565,1.08499018){\color[rgb]{0,0,0}\makebox(0,0)[t]{\lineheight{1.25}\smash{\begin{tabular}[t]{c}-13\end{tabular}}}}%
    \put(0.02852841,0.94899171){\color[rgb]{0,0,0}\rotatebox{90}{\makebox(0,0)[t]{\lineheight{1.25}\smash{\begin{tabular}[t]{c}$S_{11}$ Amplitude (dB)\end{tabular}}}}}%
    \put(0.69104085,0.98535111){\color[rgb]{0,0,0}\makebox(0,0)[t]{\lineheight{1.25}\smash{\begin{tabular}[t]{c}1296\end{tabular}}}}%
    \put(0.97845782,1.20574886){\color[rgb]{0,0,0}\makebox(0,0)[t]{\lineheight{1.25}\smash{\begin{tabular}[t]{c}(a)\end{tabular}}}}%
    \put(0.01890406,0.07017145){\color[rgb]{1,1,1}\makebox(0,0)[t]{\lineheight{1.25}\smash{\begin{tabular}[t]{c}(a)\end{tabular}}}}%
    \put(0.57001475,0.18477039){\color[rgb]{0,0,0}\makebox(0,0)[t]{\lineheight{1.25}\smash{\begin{tabular}[t]{c}Coordinate    ($\mu$m)\end{tabular}}}}%
    \put(0.7871568,0.01956377){\color[rgb]{0,0,0}\makebox(0,0)[lt]{\lineheight{1.25}\smash{\begin{tabular}[t]{l}1000  $\mu$m \end{tabular}}}}%
  \end{picture}%
\endgroup%